\documentclass[showpacs,eqsecnum,aps,prd,12pt]{revtex4}

\usepackage{graphicx}
\usepackage{revsymb}
\usepackage{amsmath}
\usepackage{amsfonts}
\usepackage{amssymb}
\usepackage{ifthen}
\usepackage{mathrsfs}

\renewcommand{\mathcal}{\mathscr}

\newcommand{\ignore}[1]{}

\newcommand{\manifold}{\mathcal}

\renewcommand{\emph}{\textsl}

\renewcommand{\d}{\mathrm{d}}
\newcommand{\diff}[2]{
    \ifthenelse{\equal{#1}{}}%
{\frac{\mathrm{d}\hphantom{#2}}{\mathrm{d}#2}}%
{\frac{\mathrm{d}#1}{\mathrm{d}#2}}}
\newcommand{\ddiff}[2]{
    \ifthenelse{\equal{#1}{}}%
{\frac{\mathrm{d}^2\hphantom{#2}}{\mathrm{d}#2^2}}
{\frac{\mathrm{d}^2#1}{\mathrm{d}#2^2}}}
\newcommand{\pardiff}[2]{
    \ifthenelse{\equal{#1}{}}%
{\frac{\partial\phantom{#2}}{\partial#2}}
{\frac{\partial#1}{\partial#2}}
}

\newcommand{\diag}{\mathop{\mathrm{diag}}}

\renewcommand{\vec}[1]{\boldsymbol{#1}}

\newcommand{\form}[1]{\boldsymbol{#1}}

\newcommand{\ip}[2]{#1\cdot#2}

\newcommand{\expct}[3][]{\langle#3|#2|#3\rangle_{\mathrm{#1}}}

\newcommand{\ket}[1]{|#1\rangle}

\newcommand{\vac}[3][]{\langle#2\rangle^{#3}_{\mathrm{#1}}}

\newcommand{\Green}[3][]{#2^{#3}_{\mathrm{#1}}}

\newcommand{\rot}{\circlearrowleft}

\begin{document}
\pagenumbering{arabic}

\title{The Renormalized Stress Tensor in {K}err Space-Time: Numerical Results
for the Hartle-Hawking Vacuum}
\author{Gavin Duffy}
\email{Gavin.Duffy@ucd.ie}
\author{Adrian C. Ottewill}
\email{Adrian.Ottewill@ucd.ie}
\affiliation{Department of
Mathematical Physics, University College Dublin, Belfield, Dublin
4, Ireland.}

\begin{abstract}
We show that the pathology which afflicts the Hartle-Hawking
vacuum on the Kerr black hole space-time can be regarded as due to
rigid rotation of the state with the horizon in the sense that
when the region outside the speed-of-light surface is removed by
introducing a mirror, there is a state with the defining features
of the Hartle-Hawking vacuum. In addition, we show that when the
field is in this state, the expectation value of the
energy-momentum stress tensor measured by an observer close to the
horizon and rigidly rotating with it corresponds to that of a
thermal distribution at the Hawking temperature rigidly rotating
with the horizon.
\end{abstract}


\maketitle

\section{Introduction}
We know that there is no Hadamard state on the Kerr black hole space-time with
the defining features of the Hartle-Hawking vacuum of respecting the isometries
of the space-time and being regular everywhere~\cite{ar:Kay91}. Frolov and
Thorne~\cite{ar:Frolov89} have shown that with certain non-standard commutation
relations a state can be found whose Feynman propagator formally has symmetry
properties necessary for regularity of the state on the outer event
horizon~\cite{ar:Hartle76}, however this state fails to be regular almost
everywhere~\cite{ar:Ottewill00a}. The concensus in the literature is that the
key property of the space-time which accounts for the non-existence of a true
Hartle-Hawking vacuum is the presence of a region in which the Killing vector
which is parallel to the null generators of the horizon becomes spacelike. In
the present article we take up a suggestion made in Ref.~\cite{ar:Frolov89} of
removing this region from the space-time by enclosing the black hole within an
axially symmetric, stationary mirror. We show that when this is done, there is
a well behaved state with the defining features of the Hartle-Hawking vacuum.
In addition, when the field is in this state the expectation value of the
energy-momentum stress tensor as measured by an observer rigidly rotating with
the horizon corresponds on the horizon to that of a thermal distribution at the
Hawking temperature rigidly rotating with the horizon.

The mirror serves to remove the superradiant normal modes, as can
be seen heuristically by noting that amplified waves which would
otherwise escape to future null infinity are reflected by it
across the future horizon. In the specific case of a mirror of
constant Boyer-Lindquist radius, we can explicitly construct a
vacuum state whose Feynman propagator has the symmetry properties
identified in Ref.~\cite{ar:Hartle76} and whose anticommutator
function does not suffer from the pathology noted in
Ref.~\cite{ar:Ottewill00a}. Although this conclusion appears to be
valid irrespective of the radius of the mirror, we generalize a
result due to Friedman~\cite{ar:Friedman78} to show that the
space-time is unstable to scalar perturbations if the mirror does
not remove all of the region outside the speed-of-light surface.
This instability is characterized by the existence of mode
solutions of the field equation with complex eigenfrequencies
which we have neglected when quantizing the field.
Kang~\cite{ar:Kang97} has shown that the contributions made by
these modes to the response function of an Unruh box are not
stationary but increase exponentially with time. We therefore
believe that when part of the region lying outside the speed-of-light
light surface is inside the mirror, it is not possible to
construct the stationary states involved in our considerations.

The layout of the paper is as follows. In Sec.~\ref{sec:FPM} we
consider the normal mode solutions of the Klein-Gordon equation
inside a mirror of constant Boyer-Lindquist radius. In
Sec.~\ref{sec:NR} we use these solutions to numerically calculate
the energy-momentum stress tensor as measured by observers rigidly
rotating with the event horizon when the radius of the mirror is
sufficiently small to remove all of the region outside the
speed-of-light surface. In Sec.~\ref{sec:SCSF} we consider the
stability of the classical scalar field in the presence of a
mirror and generalize Friedman's result. Finally, in
Sec.~\ref{sec:NSUM} we calculate numerically the eigenfrequencies
of the unstable mode solutions of the field equation present when
the black hole is inside a mirror of constant Boyer-Lindquist
radius larger than the minimum radius of the speed-of-light
surface. We follow the space-time conventions of Misner, Thorne
and Wheeler~\cite{bk:Misner73} and the notation of
Ref.~\cite{ar:Ottewill00a} throughout.

\section{Field in the Presence of a Mirror}\label{sec:FPM}
We consider the right hand region of the Kerr black hole enclosed
within a mirror $\manifold{M}$ so that the past event horizon is a
Cauchy surface for the space-time. We require that $\manifold{M}$
respects the Killing isometries of the space-time since we are
interested in a state which is invariant under these isometries.
The simplest mirror is the hypersurface $r=r_0$ since the
Klein-Gordon equation then still admits completely separable
solutions. We take the radius of the mirror to be smaller than the
minimum radius of the speed-of-light surface. We will see in
Sec.~\ref{sec:SCSF} that if the radius is larger, there are
modes of complex frequency which need to be considered in addition
to those discussed in this section. For brevity, we deal only with
the case of a field satisfying Dirichlet conditions on this hypersurface.

We can construct normal modes by
\begin{equation}
s_{\omega lm}^{\mathrm{up}}(x)=
    \begin{cases}
    \displaystyle u_{\omega lm}^{\mathrm{up}}(x)-
        \frac{R_{\omega lm}^{\mathrm{up}}(r_0)}{R_{\omega lm}^{\mathrm{in}}(r_0)}
        u_{\omega lm}^{\mathrm{in}}(x),&\omega>0,\\
    \displaystyle u_{\omega lm}^{\mathrm{up}}(x)-
        \frac{R_{\omega lm}^{\mathrm{up}}(r_0)}{R_{-\omega l-m}^{\mathrm{in}*}(r_0)}
        u_{-\omega l-m}^{\mathrm{in}*}(x),&\omega<0.
    \end{cases}
\end{equation}
By convention, the ranges of our mode labels here and in what follows are
appropriate to the ``distant observer viewpoint'' for modes with an `in' or
`out' superscript and to the ``near horizon viewpoint'' for modes with an `up'
or `down' superscript. The terminology here is that of Ref.~\cite{ar:Frolov89}.
In the ``distant observer viewpoint'' $\omega\ge0$ while in the ``near horizon
viewpoint'' $\tilde{\omega}\ge0$ where
\begin{equation}
\tilde{\omega}=\omega-m\Omega_{+}
\end{equation}
and $\Omega_{+}$ is the angular velocity of the horizon with respect to static
observers at infinity. It is straightforward to check that the modes
$s_{\omega lm}^{\mathrm{up}}$ form an orthonormal set. If we let
$h_{\omega lm}$ denote the solution of the radial equation which behaves like
$e^{i\tilde\omega r_*}$ close to the horizon then the asymptotic form of
$s_{\omega lm}^{\mathrm{up}}$ on the two horizons is
\begin{equation}
s_{\omega lm}^{\mathrm{up}}\sim
\frac{S_{\omega lm}(\cos\theta)e^{im\varphi_+}}%
{\sqrt{8\pi^2\tilde\omega(r^2+a^2)}}
    \begin{cases}
    e^{-i\tilde\omega u},&\manifold{H}^{-},\\
    \displaystyle -\frac{h_{\omega lm}(r_0)}{h_{\omega lm}^*(r_0)}
        e^{-i\tilde\omega v},&\manifold{H}^{+}.
    \end{cases}
\end{equation}
We see that, since $h_{\omega lm}(r_0)/h_{\omega lm}^*(r_0)$ is a complex
constant with unit modulus, these modes are not superradiant.

We could equally well have constructed normal modes defined by
\begin{equation}
s_{\omega lm}^{\mathrm{down}}(x)=
    \begin{cases}
    \displaystyle
    u_{\omega lm}^{\mathrm{down}}(x)-
    \frac{R_{\omega lm}^{\mathrm{down}}(r_0)}{R_{\omega lm}^{\mathrm{out}}(r_0)}
    u_{\omega lm}^{\mathrm{out}}(x),&\omega>0,\\
    \displaystyle
    u_{\omega lm}^{\mathrm{down}}(x)-
    \frac{R_{\omega lm}^{\mathrm{down}}(r_0)}%
    {R_{-\omega l-m}^{\mathrm{out}*}(r_0)}
    u_{-\omega l-m}^{\mathrm{out}*}(x),&\omega<0,
    \end{cases}
\end{equation}
which also satisfy the boundary conditions on the mirror and form an
orthonormal set. It can be shown, however, that
\begin{equation}
s_{\omega lm}^{\mathrm{down}}(x)=
    -\frac{h_{\omega lm}^*(r_0)}{h_{\omega lm}(r_0)}
    s_{\omega lm}^{\mathrm{up}}(x),
\end{equation}
and hence that these upgoing and downgoing modes differ from each
other only by a phase. It follows that the vacuum state obtain by
expanding the field in terms of either of these sets is the same
and that this state is invariant under simultaneous $t$-$\varphi$
reversal. We denote it by $\ket{B_{\manifold{M}}}$. Dropping the
now superfluous superscript from the modes, the anticommutator
function of $\ket{B_{\manifold{M}}}$ is
\begin{equation}
\Green[(1)]{G}{B_{\manifold{M}}}(x,x')=
    \sum_{l=0}^{\infty}\sum_{m=-l}^{l}\int_0^\infty\d{\tilde\omega}\,
    \Re\left[s_{\omega lm}(x)s_{\omega lm}^{*}(x')\right].
\label{eq:BmodHadamard}
\end{equation}
The modes with which the field has been expanded all have positive frequency
with respect to the Killing vector $\vec{k}_{\manifold{H}}$ given by
\begin{equation}
\vec{k}_{\manifold{H}}=\vec{k}_{\manifold{I}}+\Omega_{+}\vec{k}_{\rot}.
\end{equation}
Here, $\vec{k}_{\manifold{I}}$ and $\vec{k}_{\rot}$ are the two commuting
Killing vectors of the Kerr space-time given in Boyer-Lindquist co-ordinates by
\begin{equation}
\vec{k}_{\manifold{I}}=\partial_{t},\qquad\vec{k}_{\rot}=\partial_{\varphi}.
\end{equation}
It follows that an observer moving along an integral curve of
$\vec{k}_{\manifold{H}}$ makes measurements relative to the state
$\ket{B_{\manifold{M}}}$. Such an observer rotates rigidly with the same
velocity as the horizon with respect to static observers at infinity. We call
this observer a rigidly rotating observer (RRO).

We can extend $s_{\omega lm}$ to the left hand region of the space-time by
requiring it to be zero there. If we place a similar mirror in this region then
we can introduce a function $t_{\omega lm}$ defined by
\begin{equation}
t_{\omega lm}(U_+,V_+,\theta,\varphi_+)=
    s_{\omega lm}^{*}(-U_+,-V_+,\theta,\varphi_+),
\end{equation}
which has unit norm, satisfies the field equation, is zero in the right region
of the space-time and satisfies the boundary condition on the mirror in the
left hand region. Both of the linear combinations
\begin{align}
\sigma_{\omega lm}&=
    \frac{1}{\sqrt{1-e^{-2\pi\tilde\omega/\kappa_{+}}}}
    \left[s_{\omega lm}+
        e^{-\pi\tilde\omega/\kappa_{+}}t_{\omega lm}^{*}\right],\\
\tau_{\omega lm}&=
    \frac{1}{\sqrt{1-e^{-2\pi\tilde\omega/\kappa_{+}}}}
    \left[t_{\omega lm}+
        e^{-\pi\tilde\omega/\kappa_{+}}s_{\omega lm}^{*}\right],
\end{align}
are regular functions of $U_+$ on the past event horizon which are analytic in
the lower half of the complex $U_+$-plane and are regular functions of $V_+$ on
the future event horizon which are analytic in the lower half of the complex
$V_+$-plane. The vacuum state defined by expanding the field in terms of these
is invariant under the simultaneous $t$-$\varphi$ reversal and has a Feynman
propagator with the properties required of the Hartle-Hawking vacuum in
Ref.~\cite{ar:Hartle76}. We denote this state by $\ket{H_{\manifold{M}}}$. Its
anticommutator function is
\begin{equation}
\Green[(1)]{G}{H_{\manifold{M}}}(x,x')=
    \sum_{l=0}^{\infty}\sum_{m=-l}^{l}\int_0^\infty\d{\tilde\omega}\,
    \coth\left(\frac{\pi\tilde\omega}{\kappa_{+}}\right)
    \Re\left[s_{\omega lm}(x)s_{\omega lm}^{*}(x')\right].
\label{eq:HmodHadamard}
\end{equation}

\section{Numerical Results}\label{sec:NR}
The measurements of an RRO when the field is in the state
$\ket{H_{\manifold{M}}}$ can be calculated from
\begin{equation}
\Green[(1)]{G}{H_{\manifold{M}}}(x,x')-\Green[(1)]{G}{B_{\manifold{M}}}(x,x')=
    \sum_{l=0}^{\infty}\sum_{m=-l}^{l}\int_0^\infty\d{\tilde\omega}\,
    \frac{2}{e^{2\pi\tilde\omega/\kappa_{+}}-1}
    \Re\left[s_{\omega lm}(x)s_{\omega lm}^{*}(x')\right].
\end{equation}
The mode by mode cancellation of the high frequency divergences
which afflict both anticommutator functions in the conincident
limit makes this expression amenable to straightforward numerical
analysis. We calculated the spheroidal functions and separation
constants in essentially the same way as that outlined in
Ref.~\cite{bk:Press92}. We calculated the radial functions by
integrating Eq.~(2.6) of Ref.~\cite{ar:Ottewill00a}. The accuracy
of the integration has to be carefully considered in the case that
$M\omega$ is small and $l$ is large. The details can be found in
Ref.~\cite{th:Duffy02} and will be outlined in a later
article.

\begin{figure}[t]
\centering
\includegraphics*{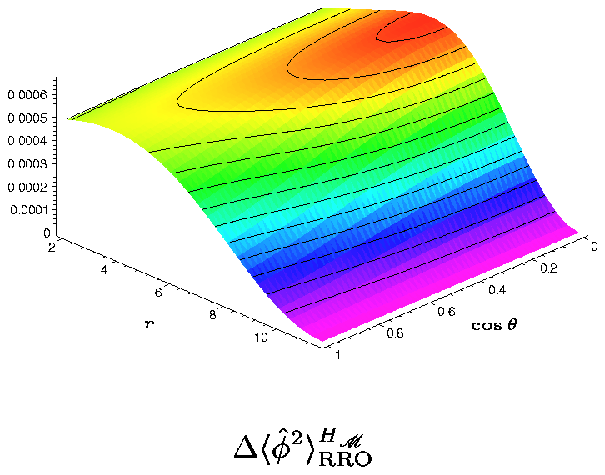}
\caption{A graph of $\vac[RRO]{\hat{\phi^2}}{H_{\manifold{M}}}$,
the expectation value of $\hat{\phi}^2$ measured by an RRO when
the field is in the state $\ket{H_{\manifold{M}}}$. The dark line
gives the value on the horizon for a thermal distribution at the
Hawking temperature rigidly rotating with the horizon. The units
are such that $M=1$.} \label{fig:phi2}
\end{figure}
\begin{figure}[p]
\centering
\includegraphics*{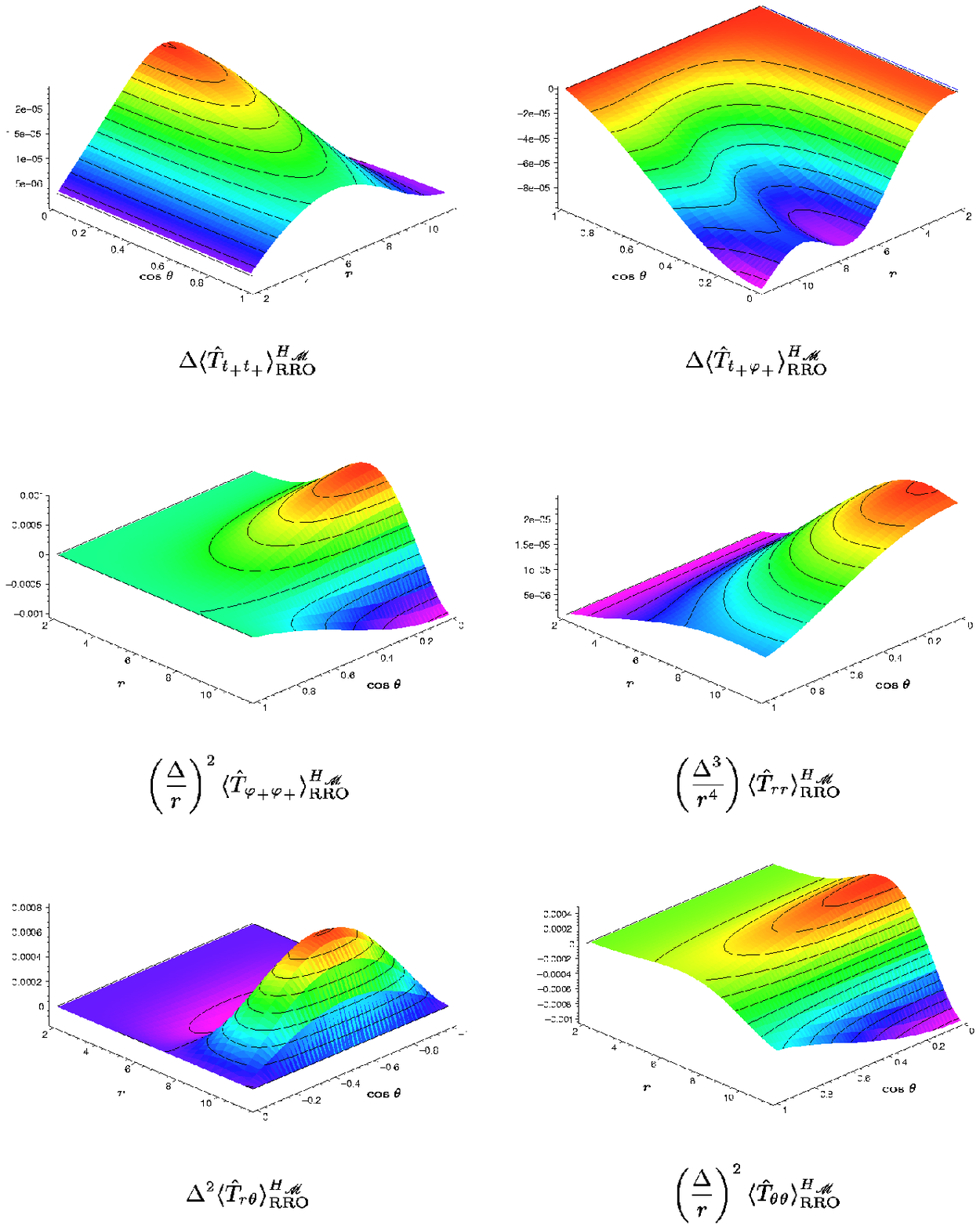}
\caption{Graphs of the components of
$\vac[RRO]{\hat{T}_{\mu\nu}}{H_{\manifold{M}}}$, the expectation value of the
energy-momentum stress tensor measured by an RRO when the field is in the state
$\ket{H_{\manifold{M}}}$. The dark line gives the value on the horizon for a
thermal distribution at the Hawking temperature rigidly rotating with the
horizon. The units are such that $M=1$.}
\label{fig:T}
\end{figure}
The graphs in Fig.~\ref{fig:phi2} and Fig.~\ref{fig:T} show the
numerically calculated values of
$\vac[RRO]{\hat{\phi^2}}{H_{\manifold{M}}}$ and
$\vac[RRO]{\hat{T}_{\mu\nu}}{H_{\manifold{M}}}$, the expectation
values measured by an RRO when the field is in the state
$\ket{H_{\manifold{M}}}$. They are given for a black hole for
which $a=0.3M$ which means that $r_{+}=1.954M$. In all of the
graphs, the units are such that $M=1$. The divergent behaviour as
the horizon is approached has been factored out and each graph
terminates on the mirror which has been given a radius
$r_{0}=11.929M$. This is just smaller than the minimum radius of
the speed-of-light surface, $r=11.935M$. See the appendix for the
details of how this is calculated. On the horizon, the components
are compared and show close agreement with those corresponding to
a thermal distribution at the Hawking temperature rigidly rotating
with the horizon. These are given by
\begin{equation}
\begin{aligned}
\vac{\phi^2}{T_H}&=\frac{T^2}{12},\\
\vac{T_{\mu\nu}}{T_H}&=\frac{T^4\pi^2}{90}
    \left[g_{\mu\nu}-
        4\frac{k_{\manifold{H}\mu}k_{\manifold{H}\nu}}%
{k_{\manifold{H}\sigma}k_{\manifold{H}}^{\sigma}}\right],
\label{eq:thermalT}
\end{aligned}
\end{equation}
where $T$ is the local temperature and is related to the Hawking temperature,
$T_H$, by
\begin{equation}
T=\frac{T_H}{\sqrt{k_{\manifold{H}\sigma}k_{\manifold{H}}^{\sigma}}},\qquad
T_H=\frac{\kappa_+}{2\pi}.
\end{equation}

\begin{figure}[t]
\centering
\includegraphics*{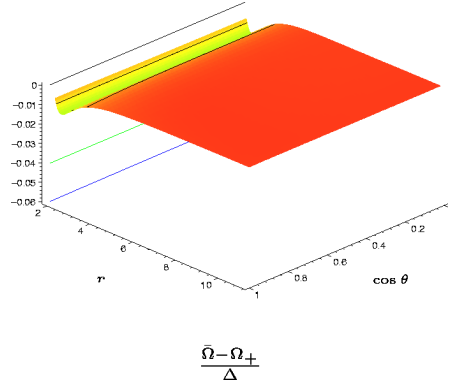}
\caption{A graph of $(\bar{\Omega}-\Omega)/\Delta$, where $\bar{\Omega}$ is the
rate of rotation of the frame in which there is no flux of energy. The black
line gives the value on the horizon for $\bar{\Omega}=\Omega_{+}$, the blue
line gives the value for $\bar{\Omega}=\Omega$ and the green line gives the
value for $\bar{\Omega}=\Omega_{\text{Carter}}$.}
\label{fig:OmegaLocal}
\end{figure}

There has been some discussion in the literature about the rate of
rotation of a Hartle-Hawking vacuum on the Kerr background, should
it be possible to define such a state. It has been suggested, for
example, that the state might appear isotropic to a ZAMO or an
observer rotating at the angular rate of the Carter tetrad. The
numerical calculations for the state $\ket{H_{\manifold{M}}}$ provide
strong evidence that it rotates rigidly up to not just leading order
but also next to leading order as the horizon is approached. This
is important because it shows that the rotation is not at the
angular rate of either a ZAMO or the Carter tetrad. One way to
see this is to consider an observer at fixed $r$ and $\theta$
rotating at an angular velocity of $\bar{\Omega}(r,\theta)$
relative to static observers at infinity. As an orthonormal tetrad
carried by the observer we can take
\begin{align}
\vec{e}_{(t)}&=\frac{\gamma}{\alpha}
\left[\partial_t+\bar{\Omega}\partial_{\varphi}\right],
\label{eq:ortho_vecs_t}\\
\vec{e}_{(\varphi)}&=\frac{1}{\gamma\tilde\Omega}\left[\partial_{\varphi}+
\frac{\gamma^2\tilde\Omega^2}{\alpha^2}(\bar{\Omega}-\Omega)
\left(\partial_t+\bar{\Omega}\partial_{\varphi}\right)\right],
\label{eq:ortho_vecs_phi}\\
\vec{e}_{(r)}&=\sqrt{\frac{\Delta}{\Sigma}}\partial_r,\\
\vec{e}_{(\theta)}&=\sqrt{\frac{1}{\Sigma}}\partial_{\theta}.
\end{align}
We have introduced here standard functions of the metric
components~\cite{bk:Misner73}
\begin{equation}
\tilde\Omega=\sqrt{g_{\varphi\varphi}},\qquad
\Omega=-\frac{g_{t\varphi}}{g_{\varphi\varphi}},\qquad
\alpha=\sqrt{\frac{g_{t\varphi}^2-
    g_{tt}g_{\varphi\varphi}}{g_{\varphi\varphi}}}\label{eq:metricFunctions}
\end{equation}
and the function
\begin{equation}
\gamma=\sqrt{\frac{g_{t\varphi}^2-g_{tt}g_{\varphi\varphi}}{g_{\varphi\varphi}
    \left|g_{tt}+2\bar{\Omega} g_{t\varphi}+
        \bar{\Omega}^2g_{\varphi\varphi}\right|}}.
\end{equation}
The function $\gamma$ is the Lorentz factor associated with the
observer's angular velocity relative to a ZAMO at the same point
and $\Omega$ is the angular velocity of this ZAMO relative to
static observers at infinity. Given an energy-momentum stress
tensor $T_{\mu\nu}$, the flux of energy in the direction of
$\vec{e}_{(\varphi)}$ measured by the observer is
\begin{align}
T_{(t\varphi)}&=\frac{1}{\alpha\tilde{\Omega}}
    \left[T_{t\varphi}+\bar{\Omega} T_{\varphi\varphi}+
        \frac{\gamma^2\tilde{\Omega}^2}{\alpha^2}(\bar{\Omega}-\Omega)
        (T_{tt}+2\bar{\Omega} T_{t\varphi}+
            \bar{\Omega}^2T_{\varphi\varphi})\right]\\
&=\frac{\gamma^2\tilde{\Omega}}{\alpha^3}
    \left[\left(\frac{\alpha^2}{\tilde{\Omega}^2}+\Omega^2\right)T_{t\varphi}-
        \Omega T_{tt}+
        \bar{\Omega}\left(T_{tt}+
            \left(\frac{\alpha^2}{\tilde{\Omega}^2}+\Omega^2\right)
                T_{\varphi\varphi}\right)+
        \bar{\Omega}^2(T_{t\varphi}+\Omega T_{\varphi\varphi})
    \right]
\end{align}
The angular velocity at which the observer must rotate in order to measure no
flux of energy is given by a root of the quadratic in $\bar{\Omega}$ appearing
in the square brackets above. From the numerical data for
$\vac[RRO]{\hat{T}_{\mu\nu}}{H_{\manifold{M}}}$, we can solve to find
$\bar{\Omega}$. Fig.~\ref{fig:OmegaLocal} gives a graph of
$(\bar{\Omega}-\Omega_{+})/\Delta$. This is compared on the horizon with the
value for an RRO, a ZAMO and an observer rotating with the same angular
velocity as the Carter tetrad. These are given by
\begin{equation}
\lim_{r\to r_{+}}
    \left\{\frac{\bar{\Omega}-\Omega_{+}}{\Delta}\right\}=
\begin{cases}
0,&\bar{\Omega}=\Omega_{+},\\
-\frac{\Omega_{+}}{4M^2r_{+}\kappa_{+}},&\bar{\Omega}=\Omega_{\text{Carter}},\\
\Omega_{+}\left[\Omega_{+}^2\sin^2\theta+
    \frac{r_{-}}{4M^2r_{+}^2\kappa_{+}}\right],&\bar{\Omega}=\Omega.
\end{cases}
\end{equation}
The graph shows close agreement with the value corresponding to an RRO and not
with either of the other two.

It might be supposed that the behaviour of
$\vac[RRO]{\hat{T}_{\mu\nu}}{H_{\manifold{M}}}$ close to the horizon is
derivable by an asymptotic analysis in the manner of Candelas, Chrzanowski and
Howard~\cite{ar:Candelas81}. Indeed, close to the horizon the contributions to
the mode sums from the $u_{\omega lm}^{\mathrm{up}}$ dominate and we find that
\begin{equation}
\vac[RRO]{\hat{T}_{\mu\nu}}{H_{\manifold{M}}}\sim
    \expct{\hat{T}_{\mu\nu}}{CCH^{-}}-\expct{\hat{T}_{\mu\nu}}{B^{-}},
    \qquad(r\to r_{+}),
\label{eq:Hasymp}
\end{equation}
the right hand side of which has been calculated in Ref.~\cite{ar:Candelas81}
for the electromagnetic field. We wish to point out, however, that this
analysis is unfortunately in error except at the poles of the horizon. The
result given in Eq.~(3.7) of Ref.~\cite{ar:Candelas81} is obtained by noting
that the dominant contributions come from the high $l$ modes and performing an
asymptotic analysis of the
${}^{\vphantom{\mathrm{up}}}_{\pm1}R_{\omega lm}^{\mathrm{up}}$ as the horizon
is approached for large $l$. The spheroidal functions are also tacitly replaced
with spherical functions. In general, however, contributions come from every
value of $m$ and since the dominant contributions to the mode sum are for
$\tilde{\omega}\approx T_{H}$, neither $\omega$ nor $m$ is necessarily small
compared to $l$. The spheroidal functions therefore cannot always be
approximated by spherical functions and the separation constant used in the
analysis of the ${}^{\vphantom{\mathrm{up}}}_{\pm1}R_{\omega lm}^{\mathrm{up}}$
cannot always be approximated by $l^2$. The poles are exceptional in that we
know that only $m=\pm1$ modes contribute there and that the analysis is
therefore valid. This can be clearly seen from the result which can be written
\begin{equation}
\expct{\hat{T}^{(\mu)}{}_{(\nu)}}{CCH^{-}}-
    \expct{\hat{T}^{(\mu)}{}_{(\nu)}}{B^{-}}\sim
    \frac{11\pi^2T^4}{45}\left(\frac{\rho^2}{2Mr_{+}}\right)\diag(-3,1,1,1),
    \qquad(r\to r_{+}),
\end{equation}
where the components are on the Carter tetrad. We have corrected
here for a typographical error in Eq.~(3.7) of
Ref.~\cite{ar:Candelas81} in which a factor of $r_{+}^2$ has been
omitted. In~\cite{ar:Casals05} it was shown numerically that in
fact
\begin{equation}
\expct{\hat{T}^{(\mu)}{}_{(\nu)}}{CCH^{-}}-
    \expct{\hat{T}^{(\mu)}{}_{(\nu)}}{B^{-}}\sim
    \frac{11\pi^2T^4}{45}\diag(-3,1,1,1),
    \qquad(r\to r_{+}),
\end{equation}
which is precisely that of a thermal distribution at the Hawking
temperature rigidly rotating with the horizon. These comments go
equally well for the scalar field; following the method of
Ref.~\cite{ar:Candelas81} we would obtain values for
$\vac[RRO]{\hat{\phi}^2}{H_{\manifold{M}}}$ and
$\vac[RRO]{\hat{T}_{\mu\nu}}{H_{\manifold{M}}}$ incorrect by the
same multiplicative factor of $\rho^2/(2Mr_{+})$.

The measurements of an RRO do not correspond to those of a thermal
distribution everywhere. It is an open question whether or not
their deviations from~(\ref{eq:thermalT}) are regular on the
horizon. It can be checked by transforming to the Kruskal
co-ordinate system that the conditions that $T_{\alpha\beta}$,
given by
\begin{equation}
T_{\alpha\beta}=T^{\mathrm{th}}_{\alpha\beta}-
    \vac[RRO]{\hat{T}_{\alpha\beta}}{H_{\manifold{M}}},
\end{equation}
is regular on both horizons are that at worst
\begin{equation}
\begin{aligned}
T_{\varphi_{+}\varphi_{+}}&=O(1),&\quad
T_{\theta\theta}&=O(1),&\quad
    T_{r\theta}&=O(1),\\
T_{t_{+}t_{+}}&=O(\Delta),&\quad
T_{t_{+}\varphi_{+}}&=O(\Delta),&\quad
    T_{t_{+}t_{+}}+\left(\frac{\Delta}{2Mr}\right)^2T_{rr}&=O(\Delta^2).
\end{aligned}
\end{equation}
These components are in the rigidly rotating co-ordinate system
$\{t_{+},r,\theta,\varphi_{+}\}$ and we have placed a subscript on
$t$ to avoid confusion with the Boyer-Lindquist co-ordinates. We
have not been able to verify these numerically. It was conjectured
by Christensen and Fulling~\cite{ar:Christensen77} that for the
Schwarzschild black hole, the measurements of a static observer
when the field is in the Hartle-Hawking state would be exactly
thermal everywhere. This was later shown not to be the case by
Jensen, McLaughlin and Ottewill~\cite{ar:Jensen92}. Although the
leading order of the deviation from isotropy is zero on the
horizon in agreement with the asymptotic analyses
of~\cite{ar:Christensen77} and~\cite{ar:Candelas80}, even in this
case it remains undetermined whether or not the total deviation is
regular there.

\section{Stability of the Classical Scalar Field}\label{sec:SCSF}
The numerical calculations of the previous section fail if the radius of the
mirror is increased to include any of the region in which
$\vec{k}_{\manifold{H}}$ becomes spacelike. If any of this region lies inside
the mirror, new modes of the field equation come into existence which are also
regular on the horizon and satisfy the boundary conditions on the mirror.
These new modes have complex eigenfrequencies and are therefore characterized
by the existence of unstable solutions of the field equation. We will show in
this section that such solutions exist for all sufficiently high $m$.
Kang~\cite{ar:Kang97} has demonstrated how to quantize the scalar field on the
Kerr background in the presence of complex frequency modes. He has shown that
the contributions made by them to the response function of an Unruh box are not
stationary but increase exponentially with time and we expect that the
anticommutator function will show similar behaviour. It follows that neither of
the stationary states $\ket{B_{\manifold{M}}}$ and $\ket{H_{\manifold{M}}}$
exists unless the region outside the speed-of-light surface is removed by
the mirror.

\begin{figure}[t]
\centering
\includegraphics*{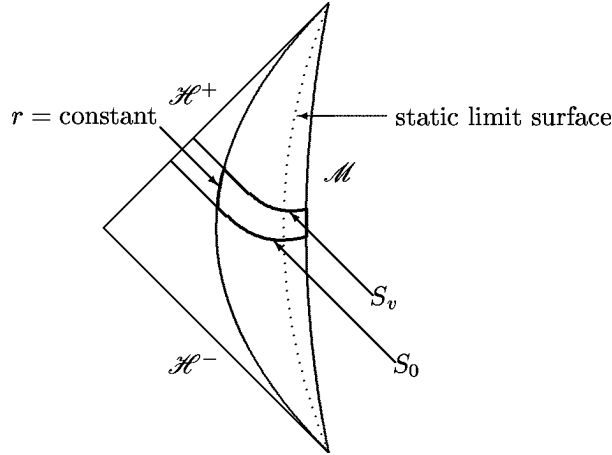}
\caption{A space-time diagram of the region bounded by the surfaces $S_{t}$,
$S_{t'}$ and the mirror.}
\label{fig:Friedman1}
\end{figure}

The analysis presented in this section is based heavily on that of
Friedman~\cite{ar:Friedman78}. The Lagrangian for the complex scalar field is
\begin{equation}
L=-\frac{1}{2}\int_{S_{t}}\form{\eta}^{(3)}
    \alpha\nabla^{\mu}\phi\nabla_{\mu}\phi^{*},
\end{equation}
where $S_{t}$ is a hypersurface of constant $t$, $\form{\eta}^{(3)}$ is the
volume $3$-form induced on $S_{t}$ by the metric and $\alpha$ is the lapse
function given in Eq.~(\ref{eq:metricFunctions}) associated with the
past pointing unit normal to $S_{t}$ which we denote by $\vec{n}$. The field
equation is
\begin{equation}
\nabla^{\mu}\nabla_{\mu}\phi=0.
\end{equation}
The total charge of the field at time $t$ is given by
\begin{equation}
\mathcal{Q}_{t}=i\int_{S_{t}}\form{\eta}^{(3)}
    n^{\mu}\left[\phi^{*}\tensor{\nabla}_{\mu}\phi\right].
\label{eq:charge}
\end{equation}
The total energy of the field at time $t$ depends on the Killing vector with
which we define derivatives with respect to time. An appropriate vector is one
of the form
\begin{equation}
\vec{k}=\vec{k}_{\mathcal{I}}+\Omega_{0}\vec{k}_{\rot},
\qquad
\text{$\Omega_{0}$ constant,}
\end{equation}
of which there is no preferred choice. The energy of the field on $S_{t}$ is
\begin{equation}
\mathcal{E}_{t}=\int_{S_{t}}\form{\eta}^{(3)}n_{\mu}E^{\mu},
\label{eq:energy}
\end{equation}
where
\begin{equation}
E^{\mu}=S^{\mu}{}_{\nu}k^{\nu},\quad
S_{\mu\nu}=\nabla_{(\mu}\phi\nabla_{\nu)}\phi^{*}-
    \frac{1}{2}g_{\mu\nu}\nabla_{\sigma}\phi\nabla^{\sigma}\phi^{*}.
\label{eq:Ecurrent}
\end{equation}
Consider the closed surface formed by the $S_{t}$, $S_{t'}$ and the mirror as
in Fig.~\ref{fig:Friedman1}.
For a solution of the field equation, $\phi^{*}\tensor{\nabla}_{\mu}\phi$ is
divergence free and so its flux over this surface is zero. Its flux over the
mirror is zero by virtue of the boundary conditions and so its flux over
$S_{t}$ is minus that over $S_{t'}$. That is, $\mathcal{Q}_{t}$ is independent
of $t$. These comments go equally well for $\vec{E}$ and so $\mathcal{E}_{t}$
is also independent of $t$.

Consider a solution of the field equation of the form
\begin{equation}
\phi_{\omega m}=\phi_{0}(r,\theta)e^{-i\omega t+im\varphi}.
\end{equation}
Suppose that $\omega$ is complex so that the solution is unstable, growing
exponentially either forwards or backwards in time. We find that
\begin{align}
\mathcal{Q}_{t}&=2e^{2\omega_{I}t}\int_{S_{t}}\form{\eta}^{(3)}
    \frac{1}{\alpha}(\omega_{R}-m\Omega)|\phi_{0}|^2,
\label{eq:modeCharge}\\
\mathcal{E}_{t}&=
    \frac{e^{2\omega_{I}t}}{2}\int_{S_{t}}\form{\eta}^{(3)}\alpha
    \left\{
        g^{rr}\left|\pardiff{\phi_0}{r}\right|^2+
        g^{\theta\theta}\left|\pardiff{\phi_0}{\theta}\right|^2+
        \frac{1}{\alpha^2}
        \left[(\omega_{R}-m\Omega_{0})^{2}+\omega_{I}^{2}-
            m^2\frac{\ip{\vec{k}}{\vec{k}}}{\tilde{\Omega}^2}\right]
        |\phi_0|^2
    \right\},
\label{eq:modeEnergy}
\end{align}
where
\begin{equation}
\omega_{R}=\Re[\omega],\qquad \omega_{I}=\Im[\omega].
\end{equation}
We know that $\mathcal{Q}_{t}$ and $\mathcal{E}_{t}$ do not depend on $t$ and
so it must be that the integrals in Eq.~(\ref{eq:modeCharge}) and
Eq.~(\ref{eq:modeEnergy}) vanish so that $\mathcal{Q}_{t}=0$ and
$\mathcal{E}_{t}=0$. Thus, for an unstable mode
\begin{align}
\int_{S_{t}}\form{\eta}^{(3)}
    \frac{1}{\alpha}(\omega_{R}-m\Omega)|\phi_{0}|^2&=0\label{eq:noCharge}\\
\int_{S_{t}}\form{\eta}^{(3)}\alpha\left\{
        g^{rr}\left|\pardiff{\phi_0}{r}\right|^2+
        g^{\theta\theta}\left|\pardiff{\phi_0}{\theta}\right|^2+
        \frac{1}{\alpha^2}
        \left[(\omega_{R}-m\Omega_{0})^{2}+\omega_{I}^{2}-
            m^2\frac{\ip{\vec{k}}{\vec{k}}}{\tilde{\Omega}^2}\right]
        |\phi_0|^2
    \right\}&=0.\label{eq:noEnergy}
\end{align}
Now if $\vec{k}$ can be chosen so that it remains timelike everywhere then it
is clear that the integral on the left hand side of Eq.~(\ref{eq:noEnergy})
must be positive for a non-trivial mode and so this condition cannot be met.
There are therefore no unstable modes in this case. In particular, putting
$\vec{k}=\vec{k}_{\mathcal{H}}$, we see that there are no unstable modes if the
radius of the mirror is smaller than the minimum radius of the speed-of-light
surface.

We could equally well have restricted the field to the region outside the
mirror. This time putting $\vec{k}=\vec{k}_{\mathcal{I}}$, we see that there
are no unstable modes if the radius of the mirror is larger than the maximum
radius of the stationary limit surface. On the other hand,
Friedman~\cite{ar:Friedman78} has shown that for a star with an ergoregion,
unstable solutions to the Klein-Gordon equation exist. It can easily be checked
that his analysis is valid for the Kerr space-time with the event horizon
removed by surrounding the black hole with a mirror. We now show that there are
likewise unstable solutions of the Klein-Gordon equation if we consider the
space-time inside a mirror surrounding the black hole but not removing all of
the region outside the speed-of-light surface.

\begin{figure}[t]
\centering
\includegraphics*{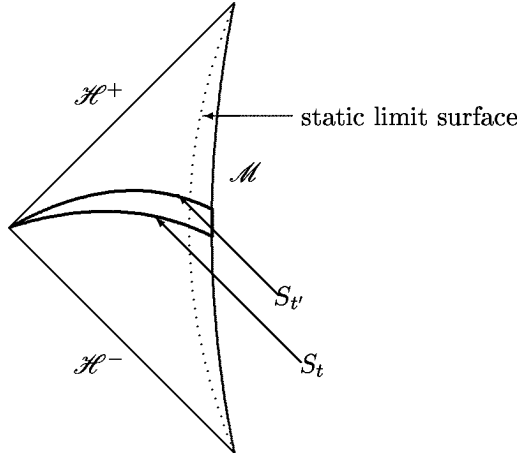}
\caption{A space-time diagram of the region bounded by the surfaces $S_{0}$,
$S_{v}$, a hypersurface of constant Boyer-Lindquist radius and the mirror.}
\label{fig:Friedman2}
\end{figure}

Let $S_{v}$ be a family of hypersurfaces related to the Killing vector
$\vec{k}_{\mathcal{H}}$ where the parameter $v$ satisfies
$k^{\mu}_{\mathcal{H}}\nabla_{\mu}v=1$. The unit normal to each of these
hypersurfaces is then given by
\begin{equation}
\vec{n}=-\alpha\,\d{v},\qquad
    \alpha=\frac{1}{\sqrt{-\nabla_{\mu}v\nabla^{\mu}v}}.
\end{equation}
The vector $\vec{k}_{\mathcal{H}}$ can be decomposed into a part which is
parallel and a part which is orthogonal to $\vec{n}$. That is,
\begin{equation}
\vec{\vec{k}_{\mathcal{H}}}=\alpha[\vec{n}+\beta\vec{m}],
\end{equation}
where
\begin{equation}
m^{\mu}m_{\mu}=1, \quad n^{\mu}m_{\mu}=0.
\end{equation}
The region inside the speed-of-light surface is clearly the region in which
$|\beta|<1$. We can write the metric as
\begin{equation}
g_{\mu\nu}=-n_{\mu}n_{\nu}+m_{\mu}m_{\nu}+j_{\mu\nu},
\end{equation}
where
\begin{equation}
    \quad j^{\mu}{}_{\nu}=\delta^{\mu}{}_{\nu}+
        4n^{[\mu}m^{\sigma]}n_{[\nu}m_{\sigma]}.
\end{equation}
The energy in the field on $S_{v}$ is
\begin{equation}
\mathcal{E}_{v}=\int_{S_{v}}\form{\eta}^{(3)}n_{\mu}E^{\mu},
\end{equation}
where $\vec{E}$ is defined in Eq.~(\ref{eq:Ecurrent}) with
$\vec{k}=\vec{k}_{\mathcal{H}}$. Suppose that close to the future horizon the
surfaces $S_{v}$ become null. Consider the region bounded by the surfaces
$S_{0}$, $S_{v}$, a hypersurface of constant Boyer-Lindquist radius $r$ and the
mirror as in Fig.~\ref{fig:Friedman2}.
Since the total flux of $\vec{E}$ over this surface is zero and its flux over
the mirror is zero by virtue of the boundary conditions, we find that
\begin{equation}
\mathcal{E}_{v}=\mathcal{E}_{0}-
    2Mr_+\int_{0}^{v}\d{v}
    \int\d{\theta}\,\d{\varphi}\,\sin\theta\,
        \left|k^{\mu}_{\mathcal{H}}\nabla_{\mu}\phi\right|^2,
\label{eq:Eevolution}
\end{equation}
In order to obtain the above expression we have taken the limit as the radius
of the hypersurface of constant $r$ tends to the radius of the event horizon
and used the condition that the field is regular on the horizon. This means
that in the advanced co-ordinate system
$\{\bar{t},r,\theta,\bar{\varphi}\}$~\cite{bk:dInverno92}, for example, $\phi$
can be expanded in $r$ about $r_+$ in terms of regular functions of $\bar{t}$,
$\theta$ and $\bar{\varphi}$. Eq.~(\ref{eq:Eevolution}) shows that a solution
which is radiating energy across the future horizon looses energy between
$S_{0}$ and $S_{v}$. To complete the proof we need to show that if
$\vec{k_{\mathcal{H}}}$ becomes spacelike somewhere within the mirror then
there is an initial configuration of the field for which $\mathcal{E}_{0}<0$.
Then Eq.~(\ref{eq:Eevolution}) implies that the field is at best marginally
unstable and will be strictly unstable unless it can settle down at late $v$ to
a non-radiative state. Furthermore, this non-radiative state state must be
time-dependent. To see this note that from the field equation,
$\mathcal{E}_{v}$ can be rewritten
\begin{equation}
\mathcal{E}_{v}=
    \frac{1}{2}\int_{S_{v}}\form{\eta}^{(3)}n^{\mu}
        \Re\left[\left(k_{\mathcal{H}}^{\sigma}\nabla_{\sigma}\phi\right)
            \tensor{\nabla}_{\mu}\phi^{*}\right]
\end{equation}
and hence is zero if $k^{\sigma}_{\mathcal{H}}\nabla_{\sigma}\phi$ is zero. We
now show that there is an initial configuration of the field for which
$\mathcal{E}<0$ if there is a region in which $\vec{k}_{\mathcal{H}}$ is
spacelike. It is straightforward to show that the energy can be rewritten
\begin{equation}
\mathcal{E}_{0}=
    \frac{1}{2}\int_{S_{0}}\form{\eta}^{(3)}\alpha\left[
    \left|n^{\mu}\nabla_{\mu}\phi\right|^2+
    2\beta\Re\left[\left(n^{\mu}\nabla_{\mu}\phi\right)
        \left(m^{\nu}\nabla_{\nu}\phi\right)\right]+
    \left|m^{\mu}\nabla_{\mu}\phi\right|^2+
    j^{\mu\nu}\nabla_{\mu}\phi\nabla_{\nu}\phi^{*}
    \right].
\end{equation}
Suppose that in the region outside the speed-of-light surface, $S_{0}$ is a
hypersurface of constant $t$. It then follows that
\begin{equation}
\vec{m}=\frac{\vec{k}_{\rot}}{\tilde\Omega},\qquad
    \beta=\frac{\tilde\Omega}{\alpha}(\Omega_{+}-\Omega).
\end{equation}
Let $O$ be an open region of $S_{0}$ lying outside the speed-of-light surface
and let $O_{R}$ be an open ball of radius $R$ contained within $O$. Let
$\varrho$ be a function which vanishes outside a compact subset of $O$, whose
value and derivatives are bounded in such a way that there is a positive
constant $K$ for which
\begin{equation}
\left|n^{\mu}\nabla_{\mu}\varrho\right|+
    \left|m^{\mu}\nabla_{\mu}\varrho\right|+
    \left|j^{\mu\nu}\nabla_{\mu}\varrho\nabla_{\nu}\varrho\right|^{1/2}<K,
\end{equation}
and which on $O_{R}$ is given by
\begin{equation}
\varrho=1.
\end{equation}
Consider the field which satisfies the initial conditions
\begin{align}
\phi&=\varrho e^{im\varphi_{+}},\\
n^{\mu}\nabla_{\mu}\phi&=-m^{\mu}\nabla_{\mu}\phi.
\end{align}
This means that
\begin{equation}
\mathcal{E}_{0}=\frac{1}{2}\int_{S_{0}}\form{\eta}^{(3)}\alpha
    \left[2(1-\beta)\left|m^{\mu}\nabla_{\mu}\phi\right|^2+
        j^{\mu\nu}\nabla_{\mu}\phi\nabla_{\nu}\phi^{*}\right]
\end{equation}
There is an $\epsilon>0$ for which $1-\beta<-\epsilon$ everywhere within $O$.
Suppose that within $O$, $\alpha$ is bounded between $\alpha_{\mathrm{min}}$
and $\alpha_{\mathrm{max}}$ and $\tilde\Omega$ is bounded above by
$\tilde\Omega_{\mathrm{max}}$. It follows that
\begin{align}
\mathcal{E}_{0}&\le
    -\epsilon\frac{\alpha_{\mathrm{min}}}{\tilde\Omega_{\mathrm{max}}^2}m^2
    \int_{O_{R}}\form{\eta}^{(3)}+
    K^2\frac{\alpha_{\mathrm{max}}}{2}\int_{O}\form{\eta}^{(3)}\\
&=-\epsilon\frac{\alpha_{\mathrm{min}}}{\tilde\Omega_{\mathrm{max}}^2}m^2
    \left|O_{R}\right|+
    K^2\frac{\alpha_{\mathrm{max}}}{2}\left|O\right|,
\end{align}
where $|\cdot|$ indicates the $3$-volume. We see that $\mathcal{E}_{0}<0$ for
all sufficiently high $m$.

\section{Numerical Search for Unstable Modes}\label{sec:NSUM}
Although we can say very little analytically about the occurence of complex
frequency modes, we can obtain a bounded region of the complex plane within
which the frequency must lie and thereby search for them numerically. From
Eq.~(\ref{eq:noCharge}) we find that
\begin{alignat}{2}
0&\le\omega_{R}\le m\Omega_{+},&\qquad m&>0,
\label{eq:realBound1}\\
0&\le\tilde{\omega}_{R}\le -m\Omega_{+},&\qquad m&<0.
\label{eq:realBound2}
\end{alignat}
Using Eq.~(\ref{eq:noCharge}), we can write Eq.~(\ref{eq:noEnergy}) as
\begin{equation}
\int_{S_{t}}\form{\eta}^{(3)}\alpha
    \left\{
        g^{rr}\left|\pardiff{\phi_0}{r}\right|^2+
        g^{\theta\theta}\left|\pardiff{\phi_0}{\theta}\right|^2+
        \frac{1}{\alpha^2}
        \left[\omega_{I}^{2}-(\omega_{R}-m\Omega)^2+
            m^2\frac{\alpha^2}{\tilde\Omega^2}\right]|\phi_0|^2
    \right\}=0.
\end{equation}
From the bound obtained for $\omega_{R}$, we find that
\begin{alignat}{2}
\omega_{I}&\le m\Omega_{+},&\qquad m&>0,
\label{eq:imagBound1}\\
\omega_{I}&\le-m\Omega_{+},&\qquad m&<0.
\label{eq:imagBound2}
\end{alignat}
Note from this that if a mode with complex frequency exists then
the real part of its frequency is in the range which we associate
with superradiance in the absence of a mirror. It is clear that
there are no axially symmetric modes with complex frequency. It is
also clear that when $a$ is set to zero and the space-time reduces
to the Schwarzschild space-time, no such modes can occur as is
well known. Curiously, the bound on the complex part of the
frequency obtained in a similar analysis given by Detweiler and
Ipser~\cite{ar:Detweiler73} does not imply this.

Specializing to the case of a field satisfying Dirichlet conditions on the
hypersurface $r=r_0$, we can now numerically search for complex frequency modes
within this region. We look for a solution of the form
\begin{equation}
\phi=\frac{1}{\sqrt{r^2+a^2}}R_{\omega lm}(r)S_{\omega lm}(\cos\theta)
    e^{-i\omega t+im\varphi},
\end{equation}
where
\begin{alignat}{2}
R_{\omega lm}&\sim e^{-i\tilde\omega r_*},&\qquad&(r_*\to-\infty),
\label{eq:Hbc}\\
R_{\omega lm}&=0,&\qquad&r=r_0.
\end{alignat}
This can be cast as a root-finding problem for a complex function of a complex
variable. The variable is $\omega$ and the function is $R_{\omega lm}(r_0)$,
obtained by integrating the radial differential equation from the horizon
subject to the condition~(\ref{eq:Hbc}). Close to the horizon, $R_{\omega lm}$
increases exponentially with increasing $r_*$ and so there are no numerical
problems in maintaining the accuracy of the integration. It is necessary to
know the separation constant $\lambda_{\omega lm}$ and we have calculated this
by using a power series expansion in $a\omega$~\cite{bk:Abramowitz64}. The
root-finding algorithm we have used is an iterative scheme known as Muller's
method~\cite{bk:Press92}.

\begin{figure}[t]
\centering
\includegraphics*{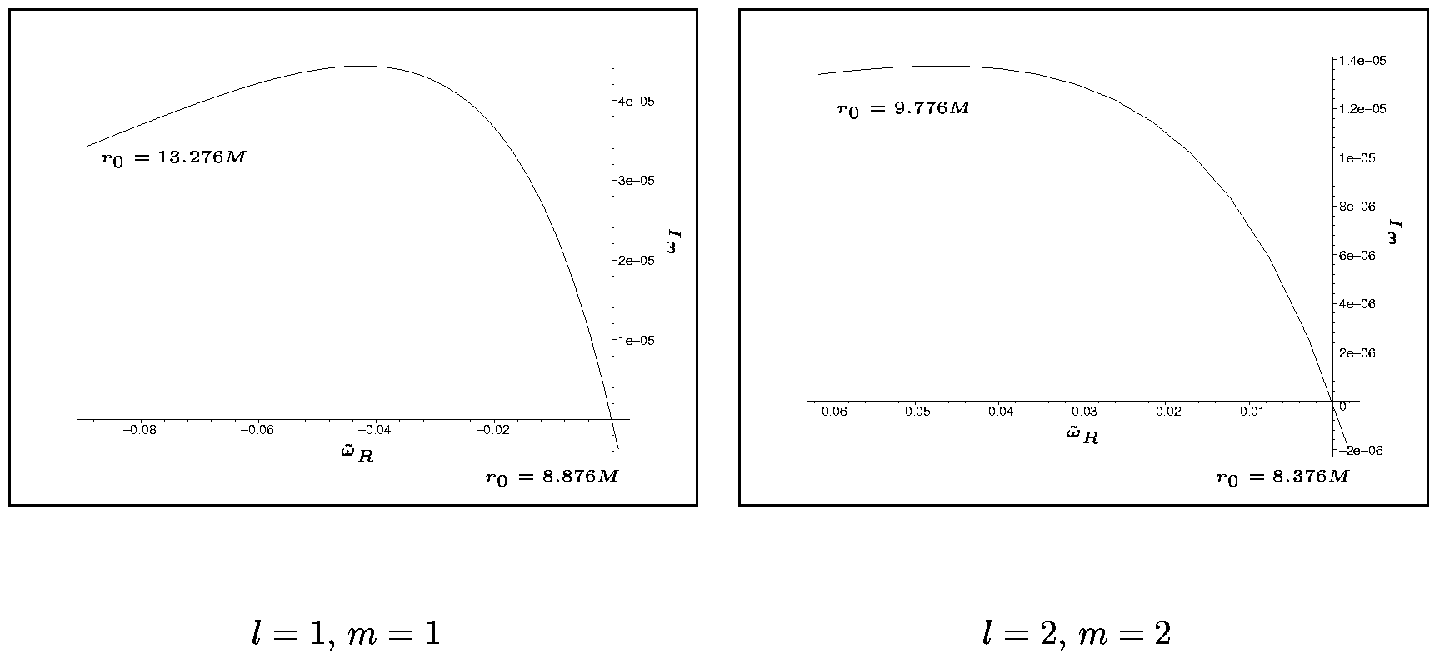}
\caption{A graph of the complex eigenfrequencies corresponding to solutions of
the Klein-Gordon equation which are null in the field theory inside a mirror of
constant Boyer-Lindquist radius $r_0$ for a black hole with $a=0.95M$. As $r_0$
decreases, the frequencies pass into the lower complex plane through the point
$\tilde\omega=0$ before the minimum radius of the speed-of-light surface,
$r_{\mathrm{SOL}}(\pi/2)=1.675855M$, is reached. The eigenfrequency is given
in units in which $M=1$.}
\label{fig:nullModes}
\end{figure}

Fig.~\ref{fig:nullModes} shows some of the results. We fix the
value of $a$ and track the movement of the frequencies found as
the value of $r_0$ changes. As the radius of the mirror decreases,
the frequencies approach the real axis and pass into the lower
half of the complex plane before the radius of the mirror reaches
the minimum radius of the speed-of-light surface. In each case,
the point at which they cross the real axis is the critical
frequency of superradiant scattering in the absence of a mirror,
$\tilde\omega=0$. This is expected in the case that the
speed-of-light surface and the static limit surface do not
cross. The reason for this is that if $\omega$ is real then $\phi$
satisfies the conditions to be an ingoing function. That is,
\begin{equation}
\phi=u_{\omega lm}^{\mathrm{in}},
\end{equation}
when suitably normalized. Since $\phi^{*}$ also vanishes on the
mirror, it is clear that the corresponding upgoing function,
$u_{\omega lm}^{\mathrm{up}}$ does also. It is straightforward to
check that $u_{\omega lm}^{\mathrm{up}}$ is an unstable mode in
the field theory outside the mirror. However, we know
from~\cite{ar:Friedman78} that there are no non-trivial modes of
this type if the mirror lies entirely outside the stationary limit
surface. The graphs in Fig.~\ref{fig:nullModes} show that this
behaviour of the complex frequencies also occurs when the black hole is
rotating sufficiently quickly that the speed-of-light surface lies
partly within the static limit surface and the argument given above
fails. The graphs are for the case $a=0.95M$, for which the relevant
surfaces are shown in Fig.~\ref{fig:SpeedOfLightSurface}.

\section{Conclusions}
In this paper, we investigated the pathology which afflicts the Hartle-Hawking
vacuum on the Kerr black hole space-time. We have done this by taking up a
suggestion made in Ref.~\cite{ar:Frolov89} of enclosing the black hole within a
mirror. The mirror serves to modify the global properties of the space-time
without altering its differential geometry. We have found that the pathology
can be regarded as due to rigid rotation of the state with the horizon in the
sense that (1) when the mirror removes the region outside the speed-of-light
surface, there is a state with the defining features of the Hartle-Hawking
vacuum; (2) when the field is in this state, the expectation value of the
energy-momentum stress tensor measured by an observer close to the horizon and
rigidly rotating with it corresponds to that of a thermal distribution at the
Hawking temperature rigidly rotating with the horizon; (3) when the mirror
encloses any part of the region outside the speed-of-light surface, the field
equation admits unstable solutions and it is not possible to construct any
stationary state such as the Hartle-Hawking vacuum. We performed the
calculation of the energy-momentum stress tensor by numerically solving for the
mode solutions of the field equation and performing mode sums. In a future
article, we will present a more detailed description of some of the numerical
techniques used. We will also give there results for the Unruh vacuum.

\appendix*

\section*{Appendix}
The speed-of-light surface is the hypersurface other than the
horizon on which $\vec{k}_{\mathcal{H}}$ becomes null. In
Boyer-Lindquist co-ordinates this is given by the condition
\begin{equation}
g_{tt}+2\Omega_+ g_{t\varphi}+\Omega_+^2 g_{\varphi\varphi}=0.
\end{equation}
In terms of $r$ and $\theta$ this can be rewritten as
\begin{equation}
\frac{\Omega_+^2\sin^2\theta}{\Sigma}(r-r_+)(r^3+Ar^2+Br+C)=0,
\end{equation}
where
\begin{align}
A&=r_+,\\
B&=a^2\cos^2\theta+2Mr_+\left(1-\frac{2Mr_+}{a^2\sin^2\theta}\right),\\
C&=(r_+-2M)\left(a^2\cos^2\theta-\frac{4M^2r_+^2}{a^2\sin^2\theta}\right)
    -4M^2r_+.
\end{align}
The cubic polynomial in $r$ which appears as a factor has only one
real root $r_{\mathrm{SOL}}$ which we can solve for in terms of
$\theta$~\cite{bk:Spiegel68},
\begin{equation}
r_{\mathrm{SOL}}(\theta)=2\sqrt{-Q}\cos\left(\frac{\Theta}{3}\right)-
    \frac{A}{3}
\end{equation}
where
\begin{equation}
Q=\frac{3B-A^2}{9},\quad P=\frac{9AB-27C-2A^3}{54},\quad
    \Theta=\cos^{-1}\left(\frac{P}{\sqrt{-Q^3}}\right).
\end{equation}
It can be checked that $r_{\mathrm{SOL}}$ has a minimum value in
the equatorial plane and that this is given by
\begin{equation}
r_{\mathrm{SOL}}\left(\frac{\pi}{2}\right)=
    \frac{r_+}{2}\left(\sqrt{1+\frac{8Mr_+}{a^2}}-1\right).
\label{eq:rSOLmin}
\end{equation}
For a black hole with $a=0.3M$, which is considered above, the minimum radius
of the speed-of-light surface is $r=11.935M$. As can bee seen, it is possible
for the speed-of-light surface to lie partly within the static limit surface.
When $a=\sqrt{2(\sqrt{2}-1)}M$ the two surfaces just touch in the equatorial
plane and in the case of an extremal black hole, $a=M$, the speed-of-light
surface just touches the outer horizon in the equatorial plane.

\begin{figure}[t]
\begin{tabular}{ccc}
\includegraphics*{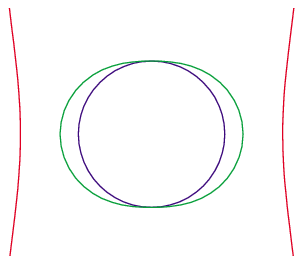}
\includegraphics*{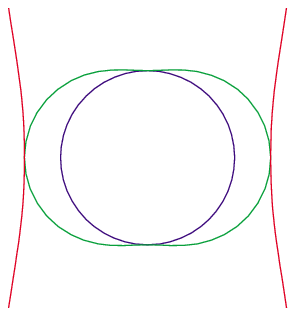}
\includegraphics*{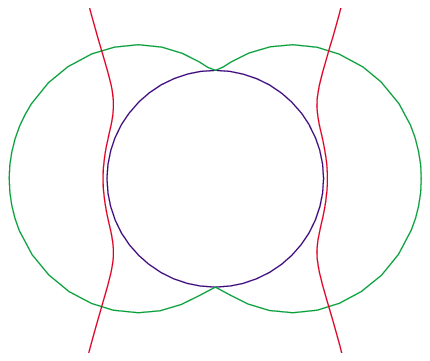}
\end{tabular}
\caption{Schematic diagrams of the horizon (blue line), static limit surface
(green line) and speed-of-light surface (red line) for $a=0.5M$,
$a=\sqrt{2(\sqrt{2}-1)}M$ and $a=0.95M$. The diagrams show a
$t=\text{constant}$, $\varphi=\text{constant}$ slice with the Boyer-Lindquist
co-ordinates $r$ and $\theta$ used as plane polar co-ordinate.}
\label{fig:SpeedOfLightSurface}
\end{figure}

\begin{acknowledgments}
We would like to thank Valery Frolov for a helpful discussion of unstable
solutions of the Klein-Gordon equation. We would also like to thank the Centre
for High-Performance Computing Applications in University College Dublin for
extensive computing time.
\end{acknowledgments}


\begin{thebibliography}{18}
\expandafter\ifx\csname natexlab\endcsname\relax\def\natexlab#1{#1}\fi
\expandafter\ifx\csname bibnamefont\endcsname\relax
  \def\bibnamefont#1{#1}\fi
\expandafter\ifx\csname bibfnamefont\endcsname\relax
  \def\bibfnamefont#1{#1}\fi
\expandafter\ifx\csname citenamefont\endcsname\relax
  \def\citenamefont#1{#1}\fi
\expandafter\ifx\csname url\endcsname\relax
  \def\url#1{\texttt{#1}}\fi
\expandafter\ifx\csname urlprefix\endcsname\relax\def\urlprefix{URL }\fi
\providecommand{\bibinfo}[2]{#2}
\providecommand{\eprint}[2][]{\url{#2}}

\bibitem[{\citenamefont{Kay and Wald}(1991)}]{ar:Kay91}
\bibinfo{author}{\bibfnamefont{B.~S.} \bibnamefont{Kay}} \bibnamefont{and}
  \bibinfo{author}{\bibfnamefont{R.~M.} \bibnamefont{Wald}},
  \bibinfo{journal}{Physics Reports} \textbf{\bibinfo{volume}{207}},
  \bibinfo{pages}{51} (\bibinfo{year}{1991}).

\bibitem[{\citenamefont{Frolov and Thorne}(1989)}]{ar:Frolov89}
\bibinfo{author}{\bibfnamefont{V.~P.} \bibnamefont{Frolov}} \bibnamefont{and}
  \bibinfo{author}{\bibfnamefont{K.~S.} \bibnamefont{Thorne}},
  \bibinfo{journal}{Phys.\ Rev.\ D} \textbf{\bibinfo{volume}{39}},
  \bibinfo{pages}{2125} (\bibinfo{year}{1989}).

\bibitem[{\citenamefont{Hartle and Hawking}(1976)}]{ar:Hartle76}
\bibinfo{author}{\bibfnamefont{J.~B.} \bibnamefont{Hartle}} \bibnamefont{and}
  \bibinfo{author}{\bibfnamefont{S.~W.} \bibnamefont{Hawking}},
  \bibinfo{journal}{Phys.\ Rev.\ D} \textbf{\bibinfo{volume}{13}},
  \bibinfo{pages}{2188} (\bibinfo{year}{1976}).

\bibitem[{\citenamefont{Ottewill and Winstanley}(2000)}]{ar:Ottewill00a}
\bibinfo{author}{\bibfnamefont{A.~C.} \bibnamefont{Ottewill}} \bibnamefont{and}
  \bibinfo{author}{\bibfnamefont{E.}~\bibnamefont{Winstanley}},
  \bibinfo{journal}{Phys.\ Rev.\ D} \textbf{\bibinfo{volume}{62}},
  \bibinfo{pages}{084018} (\bibinfo{year}{2000}).

\bibitem[{\citenamefont{Friedman}(1978)}]{ar:Friedman78}
\bibinfo{author}{\bibfnamefont{J.~L.} \bibnamefont{Friedman}},
  \bibinfo{journal}{Commun.\ Math.\ Phys} \textbf{\bibinfo{volume}{63}},
  \bibinfo{pages}{243} (\bibinfo{year}{1978}).

\bibitem[{\citenamefont{Kang}(1997)}]{ar:Kang97}
\bibinfo{author}{\bibfnamefont{G.}~\bibnamefont{Kang}},
  \bibinfo{journal}{Phys.\ Rev.\ D} \textbf{\bibinfo{volume}{55}},
  \bibinfo{pages}{7563} (\bibinfo{year}{1997}).

\bibitem[{\citenamefont{Misner et~al.}(1973)\citenamefont{Misner, Thorne, and
  Wheeler}}]{bk:Misner73}
\bibinfo{author}{\bibfnamefont{C.~W.} \bibnamefont{Misner}},
  \bibinfo{author}{\bibfnamefont{K.~S.} \bibnamefont{Thorne}},
  \bibnamefont{and} \bibinfo{author}{\bibfnamefont{J.~A.}
  \bibnamefont{Wheeler}}, \emph{\bibinfo{title}{Gravitation}}
  (\bibinfo{publisher}{W.\ H.\ Freeman and Company}, \bibinfo{address}{New
  York}, \bibinfo{year}{1973}).

\bibitem[{\citenamefont{Press et~al.}(1992)\citenamefont{Press, Teukolsky,
  Vetterling, and Flannery}}]{bk:Press92}
\bibinfo{author}{\bibfnamefont{W.~H.} \bibnamefont{Press}},
  \bibinfo{author}{\bibfnamefont{S.~A.} \bibnamefont{Teukolsky}},
  \bibinfo{author}{\bibfnamefont{W.~T.} \bibnamefont{Vetterling}},
  \bibnamefont{and} \bibinfo{author}{\bibfnamefont{B.~P.}
  \bibnamefont{Flannery}}, \emph{\bibinfo{title}{Numerical Recipes in Fortran}}
  (\bibinfo{publisher}{Cambridge University Press},
  \bibinfo{address}{Cambridge}, \bibinfo{year}{1992}), \bibinfo{edition}{2nd}
  ed.

\bibitem[{\citenamefont{Duffy}(2002)}]{th:Duffy02}
\bibinfo{author}{\bibfnamefont{G.}~\bibnamefont{Duffy}}, Ph.D. thesis,
  \bibinfo{school}{University College Dublin} (\bibinfo{year}{2002}).

\bibitem[{\citenamefont{Candelas et~al.}(1981)\citenamefont{Candelas,
  Chrzanowski, and Howard}}]{ar:Candelas81}
\bibinfo{author}{\bibfnamefont{P.}~\bibnamefont{Candelas}},
  \bibinfo{author}{\bibfnamefont{P.}~\bibnamefont{Chrzanowski}},
  \bibnamefont{and} \bibinfo{author}{\bibfnamefont{K.~W.}
  \bibnamefont{Howard}}, \bibinfo{journal}{Phys.\ Rev.\ D}
  \textbf{\bibinfo{volume}{24}}, \bibinfo{pages}{297} (\bibinfo{year}{1981}).

\bibitem[{\citenamefont{Casals and Ottewill}(2005)}]{ar:Casals05}
\bibinfo{author}{\bibfnamefont{M.}~\bibnamefont{Casals}} \bibnamefont{and}
  \bibinfo{author}{\bibfnamefont{A.~C.} \bibnamefont{Ottewill}},
  \bibinfo{journal}{Phys.\ Rev.\ D} \textbf{\bibinfo{volume}{71}},
  \bibinfo{pages}{124061} (\bibinfo{year}{2005}).

\bibitem[{\citenamefont{d'Inverno}(1992)}]{bk:dInverno92}
\bibinfo{author}{\bibfnamefont{R.}~\bibnamefont{d'Inverno}},
  \emph{\bibinfo{title}{Introducing {E}instein's Relativity}}
  (\bibinfo{publisher}{Clarendon Press}, \bibinfo{address}{Oxford, England},
  \bibinfo{year}{1992}).

\bibitem[{\citenamefont{Detweiler and Ipser}(1973)}]{ar:Detweiler73}
\bibinfo{author}{\bibfnamefont{S.~L.} \bibnamefont{Detweiler}}
  \bibnamefont{and} \bibinfo{author}{\bibfnamefont{J.}~\bibnamefont{Ipser}},
  \bibinfo{journal}{The Astrophysical Journal} \textbf{\bibinfo{volume}{185}},
  \bibinfo{pages}{675} (\bibinfo{year}{1973}).

\bibitem[{\citenamefont{Abramowitz and Stegun}(1964)}]{bk:Abramowitz64}
\bibinfo{author}{\bibfnamefont{M.}~\bibnamefont{Abramowitz}} \bibnamefont{and}
  \bibinfo{author}{\bibfnamefont{I.~A.} \bibnamefont{Stegun}},
  \emph{\bibinfo{title}{Handbook of Mathematical Functions, with Formulas,
  Graphs and Mathematical Tables}} (\bibinfo{publisher}{National Bureau of
  Standards}, \bibinfo{address}{Washington}, \bibinfo{year}{1964}).

\bibitem[{\citenamefont{Christensen and Fulling}(1977)}]{ar:Christensen77}
\bibinfo{author}{\bibfnamefont{S.~M.} \bibnamefont{Christensen}}
  \bibnamefont{and} \bibinfo{author}{\bibfnamefont{S.~A.}
  \bibnamefont{Fulling}}, \bibinfo{journal}{Phys.\ Rev.\ D}
  \textbf{\bibinfo{volume}{15}}, \bibinfo{pages}{2088} (\bibinfo{year}{1977}).

\bibitem[{\citenamefont{Jensen et~al.}(1992)\citenamefont{Jensen, McLaughlin,
  and Ottewill}}]{ar:Jensen92}
\bibinfo{author}{\bibfnamefont{B.~P.} \bibnamefont{Jensen}},
  \bibinfo{author}{\bibfnamefont{J.~G.} \bibnamefont{McLaughlin}},
  \bibnamefont{and} \bibinfo{author}{\bibfnamefont{A.~C.}
  \bibnamefont{Ottewill}}, \bibinfo{journal}{Phys.\ Rev.\ D}
  \textbf{\bibinfo{volume}{45}}, \bibinfo{pages}{3002} (\bibinfo{year}{1992}).

\bibitem[{\citenamefont{Candelas}(1980)}]{ar:Candelas80}
\bibinfo{author}{\bibfnamefont{P.}~\bibnamefont{Candelas}},
  \bibinfo{journal}{Phys.\ Rev.\ D} \textbf{\bibinfo{volume}{21}},
  \bibinfo{pages}{2185} (\bibinfo{year}{1980}).

\bibitem[{\citenamefont{Spiegel}(1968)}]{bk:Spiegel68}
\bibinfo{author}{\bibfnamefont{M.~R.} \bibnamefont{Spiegel}},
  \emph{\bibinfo{title}{Mathematical Handbook of Formulas and Tables}}
  (\bibinfo{publisher}{McGraw-Hill}, \bibinfo{address}{New York},
  \bibinfo{year}{1968}).

\end{thebibliography}

\end{document}